\documentclass[10pt, onecolumn]{article}
\usepackage{graphicx, amsmath, amssymb}
\usepackage{psfrag}
\usepackage{epstopdf}
\usepackage{authblk}
\usepackage{float}
\usepackage{hyperref}
\usepackage[a4paper,left=30mm,right=30mm,top=20mm,bottom=15mm,bindingoffset=0cm]{geometry}

\title{Geometrically nonlinear modelling of pre-stressed viscoelastic fibre-reinforced composites with application to arteries}
\author[1, 2]{I I Tagiltsev\thanks{i.i.tagiltsev@gmail.com}}
\author[1, 2]{A V Shutov\thanks{alexey.v.shutov@gmail.com}}
\affil[1]{Lavrentyev Institute of Hydrodynamics, pr. Lavrentyeva 15, 630090, Novosibirsk, Russia}
\affil[2]{Novosibirsk State University, ul. Pirogova 1, 630090, Novosibirsk, Russia}

\begin{document}
\maketitle

%_____________________________________________________________________

\begin{abstract}
Modelling of mechanical behaviour of pre-stressed fibre-reinforced composites is considered in a geometrically exact setting.
A general approach which includes two different reference configurations is employed: one configuration corresponds to the load-free state of the structure and another one to the stress-free state of each material particle.
The applicability of the approach is demonstrated in terms of a viscoelastic material model; both the matrix and the fibre are modelled using a multiplicative split of the deformation gradient tensor; a transformation rule for initial conditions is elaborated and specified.
Apart from its simplicity, an important advantage of the approach is that well-established numerical algorithms can be used for pre-stressed inelastic structures.
The interrelation between the advocated approach and the widely used ``opening angle'' approach is clarified.
A full-scale FEM simulation confirms the main predictions of the ``opening angle'' approach.
%Although the ``opening angle'' approach is based on simplified kinematics which neglects the edge effects, a full-scale FEM simulation confirms its main predictions.
A locking effect is discovered; the effect is that in some cases the opening angle of the composite is essentially smaller than the opening angles of its individual layers.
Thus, the standard cutting test typically used to analyse pre-stresses does not carry enough information and more refined experimental techniques are needed.

\end{abstract}
%_____________________________________________________________________

\section*{Nomenculature}

\begin{table}[H]
\begin{tabular}{rl}
%\hline
$\mathbf{F}$ & deformation gradient\\
$\mathbf{C}$ & right Cauchy-Green tensor\\
$\psi$ & Helmholtz  free-energy per unit mass\\
$\mathbf{\tilde{T}}$ & 2nd Piola-Kirchhoff stress tensor\\
$\mathbf{1}$ & identity tensor\\
$\mathbf{T}$ & Cauchy stress tensor\\
$\mathbf{A}^{\text{T}}$ & transpose of a tensor\\
$\text{tr}(\mathbf{A})$ & trace of a tensor\\
$\mathbf{\overline{A}}$ & unimodular part of a tensor \\
$\mathbf{A}^{\text{D}}$ & deviatoric part of a tensor\\
$K$ & current configuration\\
$\tilde{K}_\text{lf}$ & load-free reference configuration\\
$\tilde{K}_\text{sf}$ & stress-free reference configuration\\
%\hline
\end{tabular}
\end{table}

%_____________________________________________________________________

\section{Introduction}

Biological tissues, like arteries, tendons and muscles, may be considered as composites, consisting of soft isotropic matrix reinforced with embedded stiff fibres \cite{Owen2018}.
Such materials sustain large cyclic strains and show viscoelastic anisotropic mechanical behaviour.
In recent years many material models were introduced to take numerous mechanical phenomena into account:
 not only anisotropic hyperelasticity \cite{Fung1983}, \cite{Vaishnav1973}, \cite{Holzapfel2000}, \cite{Shearer2015}, \cite{Marino2018}, viscoelasticity \cite{Holzapfel2001}, \cite{Latorre2015}, \cite{Tagiltsev2018}, \cite{Liu2019} and elasto-plasticity \cite{Shutov2008}, \cite{Guan2009}, \cite{Shutov2010} were modelled, but also damage accumulation \cite{Hurschler1997}, \cite{Balzani2006}, \cite{Hamedzadeh2018} as well as growth and remodelling of living tissues \cite{Cyron2016}, \cite{Braeu2017}, \cite{Keshavarzian2018} were taken into account.
From the mechanical standpoint, the main goal of these studies is to predict the stress response of the analyzed tissue on the macroscopic level.
Mechanical stresses are not only a measure of the local load intensity, they also act as activators for a number of physiological mechanisms.
Numerous studies indicate that any reliable model of tissue growth and remodelling requires knowledge of the applied mechanical loads \cite{Keshavarzian2018}, \cite{Braeu2017}.

The primary concern of the current publication is the adequate modelling of residual stresses.
The field of residual stresses which is realized in the homeostasis is an important constituent of the overall mechanical stress field.
The residual stresses need to be taken into account since they superimpose with applied loads.
In particular, they govern the spring-back of soft tissues upon surgical manipulations like cutting and affect the propagation of the pulse wave.
In biological structures residual stresses arise due to such processes as growth, remodelling, adaptation and repair.
The presence of residual hoop stresses is indicated by the classical opening angle test \cite{Fung1983}, the residual axial stresses manifest themselves by a shortening of a blood vessel upon excising from the body \cite{Vaishnav1983}, \cite{Delfino1997}.
Some modelling approaches to residual stresses inside arteries were already considered in \cite{Holzapfel2000}, \cite{Balzani2007}, \cite{Cardamone2009} dealing with elastic (hyperelastic) stress response.
To enable accurate and efficient simulations, a combination of constitutive assumptions and numerical schemes presented in \cite{Tagiltsev2018} will be generalised here to cover the presence of pre-stresses.
In contrast to the mentioned publications, the \textit{interaction between residual stresses and viscous effects is incorporated}.
Moreover, since the mathematical structure of multiplicative plasticity models is identical to the considered equations of viscoelasticity (see, for instance, \cite{SimoMiehe1992}, \cite{Lion2000}, \cite{Shutov2019Altenbach}), the same approach can be transferred to elasto-plasticity in a straightforward way.

The so-called iso-strain approach, also known as the constrained mixture theory, is frequently used in material modelling.
It is based on the assumption that the overall stress response of the material is a sum of stress contributions provided by matrix and fibre, both of them experiencing the same deformation.
Thanks to simplified kinematics, it allows one to employ efficient numerical methods used  to assess the overall behaviour of macroscopic structures \cite{Holzapfel2000}, \cite{Gasser2002}, \cite{Tagiltsev2018}.
In this paper, in order to further reduce the computational costs, the viscoelastic properties are modelled using the Maxwell body approach (so-called ``spring-dashpot'' model).
In general, the Maxwell approach is more efficient than the use of convolution integrals \cite{Holzapfel2001}:
the Maxwell approach employs a limited number of internal variables whereas the convolution integral with a general relaxation function requires the entire deformation history.
Moreover, for the considered versions of the Maxwell bodies, efficient iteration-free time-stepping methods are already available \cite{ShutovLandgraf2013}, \cite{Shutov2018}, \cite{Tagiltsev2018}.
The applicability of the advocated approach to fibre-reinforced composites was already assessed in  \cite{Tagiltsev2018}.
Since the integration methods are non-iterational, much higher computational efficiency can be achieved, which is especially important for the analysis of clinical applications \cite{Owen2018}.

%In terms of bio-mechanics it is of crucial importance to have an accurate model, which brings relevant insights into inherent biological properties of materials.
%A blood vessel excised from the body shrinks in axial direction and springs open after being cut along \cite{Vaishnav1983}, \cite{Holzapfel2000}.
%That means that the vessel is pre-stressed both in axial and circumferential (hoop) directions.
%This paper is focused on modelling of residual stresses in viscoelastic fibre-reinforced composites.

Dealing with pre-stressed tubes, the residual hoop stresses are related to the opening angle of the tube which is cut along.
In particular, pre-stressed arteries were considered by Liu and Fung \cite{Liu1988}, \cite{Fung1989} and Holzapfel et al. \cite{Holzapfel2000} introducing vessel's stress-free configuration and it's ``opening angle'' $\alpha$ (note that the definition of the opening angle may be slightly different depending on the author).
The corresponding method will be referred to as the ``opening angle'' approach.
``Opening angle'' approach is an integral one, used for pre-stressed tubes; it includes the calculation of the stress distribution in the entire layer.
As a result, its' FEM implementation becomes overcomplicated:
every part of the body with different stress-free state requires individual modelling combined with simulation of bending into load-free state \cite{Gasser2002}.

In this work we use a more universal approach explained in Section 2.
When working with pre-stressed structures, two different reference configurations naturally appear.
First, the so-called load-free configuration corresponds to the state of the unloaded body which is characterized by the absence of external loads.
Second, in material modelling one naturally introduces the so-called stress-free configuration associated with a zero stress field.
Thus there is a need to describe the transition between these two configurations.
Stress-free and load-free configurations are connected in this study via an initial deformation gradient tensor $\mathbf{F}_0$; the $\mathbf{F}_0$ tensor may be considered as an internal variable of the material.
The use of $\mathbf{F}_0$ to adjust the pre-stresses allows us to account for various stress-free configurations for different parts of material, which might be important in a number of applications \cite{ThomasHahn1976}, \cite{Holzapfel2010}.

Whereas within the ``opening angle'' approach the initial pre-stressed state is obtained by a computationally expensive simulation of a boundary-value problem (like bending of a vessel), the $\mathbf{F}_0$-approach allows us to describe the initial state by choosing a proper $\mathbf{F}_0$-field.
In this work we demonstrate that the advocated $\mathbf{F}_0$-approach can be efficiently implemented within a non-linear FEM.
Toward that end,  constitutive equations governing the evolution and initial conditions are carefully elaborated accounting for the change of the reference configuration.
%___________________________________________________________________

In Section 3 we show that the $\mathbf{F}_0$-field can be modelled to describe vessel behaviour upon its' cutting.
We show that the $\mathbf{F}_0$-approach includes the ``opening angle'' approach as a special case.
A brief algorithm and corresponding formulae are given.
FEM-simulations of cutting a viscous pre-stressed artery are provided as a numerical test.
Section 4 contains further demonstration of the $\mathbf{F}_0$-approach;
in this section an artery consisting of two layers with different opening angles is considered and a new mechanical phenomenon is identified.
It is shown by both semi-analytical and finite element methods that opening angle of two connected layers may be essentially smaller than the minimum between the opening angles of individual constituents.
This effect is seen in the paper as a consequence of a mutual locking.
The discovered locking effect indicates that a single opening angle of the overall composite does not carry enough information for the accurate modelling of pre-stresses.
Finally, a summary of the results and concluding remarks are given in Section 5.

%_____________________________________________________________________

\section{Model of a fiber-reinforced composite with pre-stresses \label{Model}}

\begin{figure}\centering
    \psfrag{Klf}[m][][1][0]{$\tilde{K}_\text{lf}$}
    \psfrag{K}[m][][1][0]{$K$}
    \psfrag{Ksf}[m][][1][0]{$\tilde{K}_\text{sf}$}
    \psfrag{F0}[m][][1][0]{$\mathbf{F}_0$}
    \psfrag{Flf}[m][][1][0]{$\mathbf{F}^\text{lf}$}
    \psfrag{Fsf}[m][][1][0]{$\mathbf{F}^\text{sf}$}
    \scalebox{1.0}{
    \includegraphics{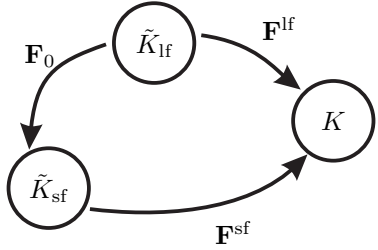}}
    \caption{General commutative diagram, showing the transition between two different reference configurations: $\tilde{K}_\text{sf}$ and $\tilde{K}_\text{lf}$. \label{GeneralComDiag}}
\end{figure}

Let $\tilde{K}_\text{lf}$ be a load-free configuration (lf-configuration) of a considered body and $\mathbf{F}^\text{lf}$ be the corresponding deformation gradient which transforms line elements from $\tilde{K}_\text{lf}$ to the current configuration $K$.
Analogously, let $\tilde{K}_\text{sf}$ be the local stress-free configuration (sf-configuration) which is characterized by zero stresses, $\mathbf{F}^\text{sf}$ is the corresponding deformation gradient.
By definition, load-free configuration is occupied by the unloaded body; it differs from the stress-free configuration due to the presence of residual stresses.
The residual stresses are caused by the incompatibility of the stress-free configuration.\footnote{The incompatibility means that the strain field which corresponds to the stress-free configuration can not be derived from a suitable field of displacements.}
The residual stresses deform the body from $\tilde{K}_\text{sf}$ to $\tilde{K}_\text{lf}$.
During a local unloading of a certain particle, its local configuration transforms from $\tilde{K}_\text{lf}$ to $\tilde{K}_\text{sf}$; the corresponding deformation gradient will be denoted as $\mathbf{F}_0$.
The stress-free deformation gradient $\mathbf{F}^\text{sf}$ is connected with the load-free deformation gradient $\mathbf{F}^\text{lf}$ by the relation
\begin{equation} \label{SF-LF}
  \mathbf{F}^\text{sf} = \mathbf{F}^\text{lf} \mathbf{F}_0^{-1}.
\end{equation}
Therefore, with a given constant $\mathbf{F}_0$ the kinematics of the material particle can be described using two different reference configurations.
A general commutative diagram is given in Figure \ref{GeneralComDiag}.
In this paper we assume that $\mathbf{F}_0$ is unimodular: $\det \mathbf{F}_0 = 1$.

\textbf{Remark 1.}
Equation \eqref{SF-LF} has a similar structure to the multiplicative split used to model growth and remodelling, see, for example, \cite{Rodriguez1994}, \cite{Braeu2017}.
In our notation, $\mathbf{F}^\text{lf}$ and $\mathbf{F}^\text{sf}$ correspond to $\mathbf{F}$ and $\mathbf{F}^\text{e}$ from \cite{Braeu2017}. In contrast to \cite{Braeu2017}, inelastic stress response is included in the current study.

%%%%%%%%%%%%%%%%%%%%%%%%%%%%%%%%%%%%%%%%%%%%%%%%%%%%%%%%%%%%%%%%%%%%%%%%%%%%%%
% REFRASE %%%%%%%%%%%%%%%%%%%%%%%%%%%%%%%%%%%%%%%%%%%%%%%%%%%%%%%%%%%%%%%%%%%%
%%%%%%%%%%%%%%%%%%%%%%%%%%%%%%%%%%%%%%%%%%%%%%%%%%%%%%%%%%%%%%%%%%%%%%%%%%%%%%

\begin{figure}\centering
    \psfrag{Klf}[m][][1][0]{$\tilde{K}_\text{lf}$}
    \psfrag{K}[m][][1][0]{$K$}
    \psfrag{Ksf}[m][][1][0]{$\tilde{K}_\text{sf}$}
    \psfrag{F0}[m][][1][0]{$\mathbf{F}_0$}
    \psfrag{Flf}[m][][1][0]{$\mathbf{F}^\text{lf}$}
    \psfrag{Fsf}[m][][1][0]{$\mathbf{F}^\text{sf}$}
    \psfrag{Ki}[m][][1][0]{$\hat{K}$}
    \psfrag{Fi}[m][][1][0]{$\mathbf{F}_\text{i}^\text{lf}$}
    \psfrag{Fisf}[m][][1][0]{$\mathbf{F}_\text{i}^\text{sf}$}
    \psfrag{Fe}[m][][1][0]{$\mathbf{\hat{F}}_\text{e}$}
    \scalebox{1.0}{
    \includegraphics{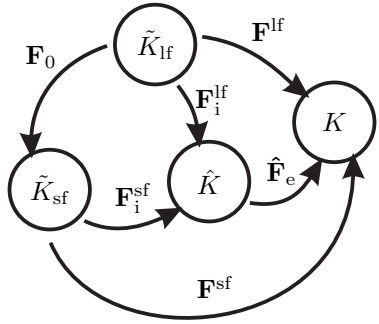}}
    \caption{Commutative diagram for multiplicative decomposition of deformation gradient. \label{ViscComDiag}}
\end{figure}

Special constitutive assumptions are essential to account for inelastic properties of material.
For these purposes we use the Sidoroff multiplicative decomposition of deformation gradient into elastic and inelastic parts.
This decomposition gives rise to the intermediate configuration $\hat{K}$, which is achieved by instantaneous unloading from the current configuration; corresponding deformation is denoted as $ \mathbf{\hat{F}}_\text{e}^{-1}$.
Respective to the stress-free configuration we have $\mathbf{F}_\text{sf} = \mathbf{\hat{F}}_\text{e} \mathbf{F}_\text{i}^\text{sf}$.
%Using the multiplicative decomposition of the deformation gradient $\mathbf{F}_\text{sf} = \mathbf{\hat{F}}_\text{e} \mathbf{F}_\text{i}^\text{sf}$ for isotropic and fibre-like Maxwell bodies yields the so-called intermediate configuration $\hat{K}$.
% At the initial
Then \eqref{SF-LF} yields the following:
\begin{equation}
  \mathbf{F}^\text{lf} = \mathbf{F}^\text{sf} \mathbf{F}_0 = \mathbf{\hat{F}}_\text{e} \mathbf{F}_\text{i}^\text{sf} \mathbf{F}_0 = \mathbf{\hat{F}}_\text{e} \mathbf{F}_\text{i}^\text{lf}, \quad \mathbf{F}_\text{i}^\text{lf} = \mathbf{F}_\text{i}^\text{sf} \mathbf{F}_0.
\end{equation}
This means that the elastic part of the deformation gradient does not depend on the choice of the reference configuration.
The specific commutative diagram is shown in Figure \ref{ViscComDiag}.

\begin{figure}\centering
    \psfrag{Klf}[m][][1][0]{$\tilde{K}_\text{lf} = K$}
    \psfrag{Ksf}[m][][1][0]{$\tilde{K}_\text{sf} = \hat{K}$}
    \psfrag{F0}[m][][1][0]{$\mathbf{F}_0$}
    \psfrag{Fi}[m][][1][0]{$\mathbf{F}_\text{i}^\text{lf}|_{t=0}$}
    \psfrag{Fe}[m][][1][0]{$\mathbf{\hat{F}}_\text{e}|_{t=0}$}
    \scalebox{1.0}{
    \includegraphics{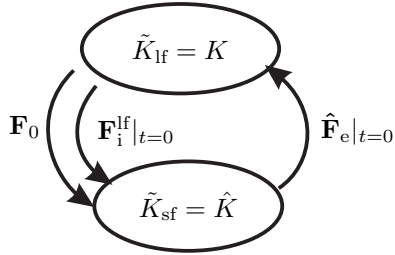}}
    \caption{Model kinematics for $ t = 0$ to clarify the nature of $\mathbf{F}_0$. \label{F0_EI}}
\end{figure}

%The mechanical interpretation of $\mathbf{F}_0$ requires thoughtful consideration here.
%On the one side, it is responsible for residual stresses and associated with elastic behaviour of a body.
%On the other side, it is included into inelastic operator $\mathbf{F}_i^{lf} = \mathbf{F}_i^{sf} \mathbf{F}_0$ and thus may be associated with inelastic behaviour.
To clarify the nature of $\mathbf{F}_0$, let us consider the following situation.
Let $\tilde{K}_\text{lf}$ be a reference configuration.
Assume that it coincides with the current configuration $K$ at the initial instance of time $t=0$: $\tilde{K}_\text{lf} = K|_{t=0}$.
Instantaneous unloading of the material particle is described by $\mathbf{\hat{F}}^{-1}_\text{e}$: the operator $\mathbf{\hat{F}}^{-1}_\text{e}$ transforms $K$ to $\hat{K}$ and brings $\tilde{K}_\text{lf}$ to $\tilde{K}_\text{sf}$.
Thus $\tilde{K}_\text{sf} = \hat{K}|_{t=0}$ in that case.
Since $\mathbf{F}^\text{lf}|_{t=0} = \mathbf{1}$ we obtain $\mathbf{F}_0 = \mathbf{F}_\text{i}^\text{lf}|_{t=0} = (\mathbf{\hat{F}}_\text{e}|_{t=0})^{-1}$.
This special case is summarized in Figure \ref{F0_EI}.
Two different interpretations of $\mathbf{F}_0$ are thus possible: it can be the initial inelastic deformation or the inverse of the initial elastic deformation.

The entire problem of modelling the initial stresses boils down to an accurate manipulation with reference configurations and initial conditions.
In the current section, a procedure for obtaining the Cauchy stress tensor as a function of the deformation gradient ${\mathbf{F}^\text{lf}}$ with respect to the load-free configuration is suggested.
%a demonstration problem is considered.
The procedure is exemplified using the model of fibre-reinforced viscoelastic composite which was advocated in \cite{Tagiltsev2018}.
That model includes Mooney-Rivlin and Holzapfel potentials associated with isotropic and fibre-anisotropic hyperelastic behaviour.
Isotropic and fibre-like Maxwell bodies are incorporated to take viscoelasticity into account.

%_____________________________________________________________________

\subsection{Isotropic and fibre-like hyperelasticity with pre-stresses}

The mechanical properties of hyperelastic materials are uniquely defined by the free energy density function.
In many cases, the natural choice of the reference configuration is $\tilde{K} = \tilde{K}_\text{sf}$, since it provides zero stresses in undeformed state.

\textbf{Isotropic hyperelasticity.}
Here we consider the two-term Mooney-Rivlin material.
By definition, the right Cauchy-Green tensor with respect to the sf-configuration $\tilde{K}_\text{sf}$ is given by
\begin{equation} \label{Csf}
  \mathbf{C}^\text{sf} = {\mathbf{F}^\text{sf}}^\text{T} {\mathbf{F}^\text{sf}}.
\end{equation}
The corresponding free energy per unit mass reads
\begin{equation} \label{MooneyRivlin}
  \rho_\text{R} \Psi_\text{MR}(\mathbf{C}^\text{sf}) = \frac{\displaystyle c_{1}}{2}(\text{tr} \overline{\mathbf{C}^\text{sf}} -3) + \frac{\displaystyle c_{2}}{2}(\text{tr}\overline{{\mathbf{C}^\text{sf}}^{-1}} - 3),
\end{equation}
where $c_{1}$ and $c_{2}$ are shear moduli of the material, $\rho_\text{R}$ is the mass density in both reference configurations, $\overline{\mathbf{C}} = (\det{\mathbf{C}})^{-1/3}\mathbf{C}$ stands for the unimodular part of ${\mathbf{C}}$.
The second Piola-Kirchhoff stress operating on the sf-configuration is computed through
\begin{equation}
	\mathbf{\tilde{T}}^\text{sf}_\text{MR} = 2\rho_\text{R} \frac{\partial \Psi_\text{MR}(\mathbf{C}^\text{sf})}{\partial \mathbf{C}^\text{sf}} = {\mathbf{C}^\text{sf}}^{-1} (c_1 \overline{\mathbf{C}^\text{sf}} - c_2 \overline{{\mathbf{C}^\text{sf}}^{-1}})^{\text{D}}.
\end{equation}
Here $\mathbf{A}^{\text{D}} $ stands for deviatoric part of a tensor.

Recalling that $\mathbf{F}^\text{lf} = \mathbf{F}^\text{sf} \mathbf{F}_{0}$ we obtain the following transformation rules
\begin{equation}\label{CsfClf}
	\mathbf{C}^\text{lf} = \mathbf{F}^\text{T}_0 \mathbf{C}^\text{sf} \mathbf{F}_{0}, \quad \mathbf{C}^\text{sf} = \mathbf{F}^{-\text{T}}_0 \mathbf{C}^\text{lf} \mathbf{F}_{0}^{-1}.
\end{equation}
The second Piola-Kirchhoff operating on the lf-configuration thus reads
\begin{equation}\label{PiolaStresses}
	\mathbf{\tilde{T}}^\text{lf}_\text{MR} = 2\rho_\text{R} \frac{\partial \Psi_\text{MR}( \mathbf{F}^{-\text{T}}_0 \mathbf{C}^\text{lf} \mathbf{F}_{0}^{-1} )}{\partial \mathbf{C}^\text{lf}} = 2\rho_\text{R} \mathbf{F}_0^{-1} \frac{\partial \Psi_\text{MR}(\mathbf{C}^\text{sf})}{\partial \mathbf{C}^\text{sf}} \mathbf{F}_0^{-\text{T}} = \mathbf{F}_0^{-1} \mathbf{\tilde{T}}^\text{sf}_\text{MR} \mathbf{F}_0^{-\text{T}}.
\end{equation}
Obviously, stresses are zero whenever $\mathbf{C}^\text{sf} = \mathbf{1}$.

\textbf{Fibre-like hyperelasticity.}
In order to model hyperelastic properties of a fibre family we consider the well-known Holzapfel potential \cite{Holzapfel2000}.
Now it is a function of $\mathbf{C}^\text{sf}$:
\begin{equation} \label{HolzapfelAniso}
  \begin{array}{l}
      \displaystyle \ \Psi_\text{Holzapfel}(\lambda ^{2}) = \frac{k_{1}}{2 k_{2}} (e^{k_{2}(\lambda^{2}-1)^{2}} - 1), \quad
      f (\lambda ^{2}) = 2k_{1}(\lambda^{2} - 1)e^{ k_{2}(\lambda^{2} - 1)^{2} },
  \end{array}
\end{equation}
where $f (\lambda ^{2}) := \frac{\displaystyle d \ \Psi_\text{Holzapfel}(\lambda ^{2}) }{\displaystyle d ( \lambda ^{2} )}$, $\lambda^{2} = \overline{\mathbf{C}^\text{sf}}:( \tilde{\mathbf{a}} \otimes \tilde{\mathbf{a}}) = \tilde{\mathbf{a}} \cdot \overline{\mathbf{C}^\text{sf}} \cdot \tilde{\mathbf{a}}$; $\tilde{\mathbf{a}}$ stands for the unit vector corresponding to the direction of fibre family in the sf-configuration, $\lambda$ is the stretch of that family (the dependence of the stretch on the choice of the sf-configuration is understood but omitted for brevity of notation),  $k_{1} > 0$ is a stress-like material parameter and $k_{2} > 0$ is a dimensionless parameter.

The second Piola-Kirchhoff stress operating on the sf-configuration here is obtained by the chain rule
\begin{equation} \label{HolzapfelPiola}
  \mathbf{\tilde{T}}^\text{sf}_\text{Holzapfel} = 2 \rho_\text{R}
  \frac{\displaystyle d \ \Psi_\text{Holzapfel}(\lambda ^{2}) }{\displaystyle d ( \lambda ^{2} )}
  \frac{\displaystyle d ( \overline{\mathbf{C}^\text{sf}}: \mathbf{M} ) }{\displaystyle d \mathbf{C}^\text{sf}}
  = 2 \rho_\text{R} f (\lambda ^{2}) \mathbb{P}_{\mathbf{C}^\text{sf}}:\mathbf{M},
\end{equation}
where $\mathbb{P}_{\mathbf{C}^\text{sf}}:\mathbf{X} =  \mathbf{X} - \frac{1}{3}\text{tr}(\mathbf{C}^\text{sf}\mathbf{X}){\mathbf{C}^\text{sf}}^{-1}$ and $\mathbf{M} = \tilde{\mathbf{a}} \otimes \tilde{\mathbf{a}}$ stands for the structural tensor of the fibre family.
Note that $\mathbf{M}$ operates on the sf-configuration.

In case of fibres, similar relations to \eqref{PiolaStresses} govern the transformation of the second Piola-Kirchhoff stresses:
\begin{multline}
	\mathbf{\tilde{T}}^\text{lf}_\text{Holzapfel} = 2\rho_\text{R} \frac{\partial \Psi_\text{Holzapfel}(  \mathbf{F}^{-\text{T}}_0 \mathbf{C}^\text{lf} \mathbf{F}_{0}^{-1} )}{\partial \mathbf{C}^\text{lf}} = \\ 2\rho_\text{R} \mathbf{F}_0^{-1} \frac{\partial \Psi_\text{Holzapfel}(\mathbf{C}^\text{sf})}{\partial \mathbf{C}^\text{sf}} \mathbf{F}_0^{-\text{T}} = \mathbf{F}_0^{-1} \mathbf{\tilde{T}}^\text{sf}_\text{Holzapfel} \mathbf{F}_0^{-\text{T}}.
\end{multline}

%_____________________________________________________________________

\subsection{Isotropic and fibre-like Maxwell bodies with pre-stresses \label{Initials}}

In the composite model advocated in \cite{Tagiltsev2018} viscoelastic material models are based on the Sidoroff multiplicative decomposition of deformation gradient tensor $\mathbf{F}$ into the elastic part $\hat{\mathbf{F}}_\text{e}$ and the inelastic part $\mathbf{F}_\text{i}$
\begin{equation}\label{MultDecSidoroff}
  \mathbf{F} = \hat{\mathbf{F}}_\text{e} \mathbf{F}_\text{i}.
\end{equation}
As already discussed, for $ \mathbf{F}^\text{sf} $ this relation takes the form $ \mathbf{F}^\text{sf} = \hat{\mathbf{F}}_\text{e} \mathbf{F}_\text{i}^\text{sf}$.
This decomposition gives rise to the inelastic right Cauchy-Green tensor operating on the sf-configuration
\begin{equation}
	\mathbf{C}_\text{i}^\text{sf} = {\mathbf{F}_\text{i}^\text{sf}}^\text{T} \mathbf{F}_\text{i}^\text{sf}.
\end{equation}

\textbf{Isotropic Maxwell body.}
The constitutive assumptions are taken from the work of Simo \& Miehe \cite{SimoMiehe1992}.
The representation on the reference configuration follows the paper \cite{Lion1997}.
The elastic properties of the isotropic Maxwell body are described with the neo-Hookean potential. On the sf-configuration we have
\begin{equation}\label{MaxIso}
  \begin{array}{l}
      \displaystyle \Psi_\text{neo-Hooke} = \Psi_\text{neo-Hooke}(\mathbf{C}^\text{sf}{\mathbf{C}^\text{sf}_\text{i}}^{-1}) = \frac{\mu}{2\rho_\text{R}} (\text{tr}\big (\overline{\mathbf{C}^\text{sf}{\mathbf{C}^\text{sf}_\text{i}}^{-1}} \big) -3),\\

      \displaystyle \tilde{\mathbf{T}}^\text{sf}_\text{neo-Hooke} = 2 \rho_\text{R} \frac{\partial \Psi_\text{neo-Hooke}( \mathbf{C}^\text{sf}{\mathbf{C}^\text{sf}_\text{i}}^{-1} ) }{\partial \mathbf{C}^\text{sf}} \mid_{\mathbf{C}^\text{sf}_\text{i} = \text{const}} = \mu {\mathbf{C}^\text{sf}}^{-1} \big(\overline{\mathbf{C}^\text{sf}}{\mathbf{C}^\text{sf}_\text{i}}^{-1} \big)^{\text{D}}.
  \end{array}
\end{equation}
The evolution equation and the initial condition for $\mathbf{C}^\text{sf}_\text{i}$ take the form
\begin{equation}\label{CinEvol}
      \displaystyle \dot{\mathbf{C}}^\text{sf}_\text{i} = \frac{1}{\eta}(\mathbf{C^\text{sf}}\tilde{\mathbf{T}^\text{sf}})^{\text{D}}\mathbf{C}^\text{sf}_\text{i} = \frac{\mu}{\eta}(\overline{\mathbf{C}^\text{sf}}{\mathbf{C}^\text{sf}_\text{i}}^{-1})^{\text{D}}\mathbf{C}_\text{i}^\text{sf}, \quad \mathbf{C}^\text{sf}_\text{i}\mid_{t = 0} = \mathbf{C}^{0}_\text{i}.
\end{equation}
Here $\mu$ is the shear modulus of the material, $ \eta$ is the material viscosity.

For the second Piola-Kirchhoff stress the transformation relations similar to \eqref{PiolaStresses} are valid
\begin{equation}
	\mathbf{\tilde{T}}^\text{lf}_\text{neo-Hooke} = \mathbf{F}_0^{-1} \mathbf{\tilde{T}}^\text{sf}_\text{neo-Hooke} \mathbf{F}_0^{-\text{T}}.
\end{equation}

An efficient implicit non-iterative time-stepping scheme for this model was suggested in \cite{ShutovLandgraf2013}. An extended model employing the Mooney-Rivlin potential can be treated using the algorithm suggested in \cite{Shutov2018}.

\textbf{Fibre-like Maxwell body.}
We make a non-restrictive assumption that the fibres are not rotated by the inelastic deformation:  $\mathbf{F}^\text{sf}_{\text{i}} \tilde{\mathbf{a}} = \lambda_{\text{i}} \tilde{\mathbf{a}} $.
Here $\lambda_\text{i}$ is the inelastic stretch of the fibre seen from the sf-configuration.
This assumption allows us to define the elastic stretch as $ \lambda_{\text{e}} = \| \mathbf{F}_{\text{e}} \tilde{\mathbf{a}} \|$.
The Sidoroff decomposition naturally yields the following split for stretches: $\lambda = \lambda_{\text{e}} \lambda_{\text{i}}$.
The model of fibre-like Maxwell body employs the aforementioned Holzapfel energy storage function \eqref{HolzapfelAniso} operating with an elastic stretch of fibre:
\begin{equation}
	\Psi_\text{viscFibre}(\lambda_{\text{e}}^{2}) = \frac{\displaystyle k_{1}}{\displaystyle  k_{2}} ( e^{ k_{2}(\lambda_{\text{e}}^{2} - 1)^{2} } - 1), \quad
      f = 2k_{1}(\lambda_{\text{e}}^{2} - 1)e^{ k_{2}(\lambda_{\text{e}}^{2} - 1)^{2} }.
\end{equation}
For the second Piola-Kirchhoff stress on the sf-configuration we obtain
\begin{equation}
      \tilde{\mathbf{T}}_\text{viscFibre}^\text{sf} = 2 \rho_\text{R}
  \frac{\displaystyle d \ \Psi_\text{viscFibre}(\lambda^{2}_\text{e}) }{\displaystyle d ( \lambda ^{2}_\text{e} )}
  \frac{\displaystyle d (  \lambda ^{2}_\text{e} ) }{\displaystyle d \mathbf{C}^\text{sf}}
  = 2 \rho_\text{R} \frac{ f (\lambda^2_\text{e})}{ \lambda^{2}_{\text{i}}} \mathbb{P}_{\mathbf{C}^\text{sf}}:\mathbf{M}
\end{equation}
while the evolution equation and the initial condition for inelastic stretch take the form:
\begin{equation}\label{lambdaEvol}
  \begin{array}{l}
      \frac{\displaystyle \dot{\lambda_{\text{i}}}}{ \displaystyle \lambda_{\text{i}} } = \frac{1}{\displaystyle\eta} f\Big( \big(\frac{\displaystyle \lambda}{\displaystyle \lambda_{\text{i}}}\big)^{2}\Big) \cdot \frac{\displaystyle \lambda^{2}}{\displaystyle \lambda^{2}_{\text{i}}} \rho_\text{R}, \quad \lambda_\text{i} \mid_{t = 0} = \lambda_\text{i}^{0}.
  \end{array}
\end{equation}
The transition between the two reference configurations is governed by
\begin{equation}
	\mathbf{\tilde{T}}^\text{lf}_\text{viscFibre} = \mathbf{F}_0^{-1} \mathbf{\tilde{T}}^\text{sf}_\text{viscFibre} \mathbf{F}_0^{-\text{T}}.
\end{equation}
An efficient algorithm for the fibre-like Maxwell body was suggested and tested in \cite{Tagiltsev2018}, \cite{ShutovTagiltsev2019}. Its extension to elasto-viscoplasticity was analysed in \cite{Shutov2019Altenbach}.
An efficient numerics is especially important when a big number of fibre families is considered for a greater accuracy \cite{Jin2016}.

\textbf{Initial conditions.}
In case of a pre-stressed material the initial conditions should be chosen in a proper way.
As is seen from \eqref{CinEvol} and \eqref{lambdaEvol} the required initial values are the inelastic right Cauchy-Green tensor $\mathbf{C}^{0}_\text{i}$ (corresponding to the isotropic Maxwell body) and the inelastic stretch $\lambda_\text{i}^{0}$ (corresponding to the fibre-like Maxwell body).
We assume that the viscous components are in their relaxed state in the lf-configuration, which yields $\mathbf{C}^{0}_\text{e} = \mathbf{1}$ and $\lambda_\text{e}^0 = 1$.
This is equivalent to $\mathbf{C}^{0}_\text{i} = \mathbf{C}^\text{sf} \mid_{t = 0}$ and $\lambda_\text{i}^{0} = \lambda\mid_{t = 0}$.

%_____________________________________________________________________

\section{Interrelation between the $\mathbf{F}_0$-approach and the ``opening angle'' approach \label{Interrelation}}

Consider a special case when the analysed body is a thick-walled tube.
In \cite{Holzapfel2000} a link between $\tilde{K}_\text{sf}$ and $\tilde{K}_\text{lf}$ is considered as a deformation $\mathbf{\chi}_\text{res}$ between the open sector of the tube and the closed tube.
The body in the sf-configuration is described in terms of cylindrical coordinates by
\begin{equation}
	R_\text{i} < R < R_\text{o}, \quad 0 < \Theta < (2 \pi - \alpha), \quad 0 < Z < L,
\end{equation}
where $R_\text{i}, R_\text{o}, \alpha$ and $L$ denote the inner and outer radii, the opening angle and the length of unstressed tube, respectively.
The geometry of the lf-configuration of the tube is defined by
\begin{equation}
	r_\text{i} < r < r_\text{o}, \quad 0 < \theta < 2 \pi, \quad 0 < z < l,
\end{equation}
where $r_\text{i}, r_\text{o}$ and $l$ denote the inner and the outer radii and the length of the tube.
Effectively, the deformation $\mathbf{\chi}_\text{res}$ coincides with $\mathbf{F}_0^{-1}$.
Now our goal is to specify the form of $\mathbf{F}_0$ for that case.

\subsection{Procedure for setting $\mathbf{F}_0$ \label{SemiAnalyt}}

The geometry relations for the reference configurations in the ``opening angle'' approach are described via
\begin{equation}
	r = \sqrt{ \frac{ R^2 - R_\text{i}^2 }{k \lambda_z} + r_\text{i}^2 }, \quad
	\theta = k \Theta, \quad
	z = \lambda_z Z,
\end{equation}
where $k = 2\pi / (2\pi - \alpha)$.
In terms of cylindrical coordinates with the orthonormal basis of $\{ \mathbf{e}_r, \mathbf{e}_\theta, \mathbf{e}_z \}$ the deformation gradient $\mathbf{F}_0$ which brings the lf-configuration to the sf-configuration takes the matrix form
\begin{equation}\label{CalibrField}
	\mathbf{F}_{0} = \begin{pmatrix}
                c f & 0 & 0 \\
                0 & \displaystyle \frac{1}{f} & 0 \\
                0 & 0 & \displaystyle \frac{1}{c}
            \end{pmatrix},
\end{equation}
where $c = {l}/{L}$ is the length ratio and the function $f$ describes the distribution of strains.
Given the incompressibility condition, $f$ can be written in both reference configurations:
\begin{equation}
	f_\text{lf}(r) = k \frac{ r }{ \sqrt{(r^2 - r_\text{i}^2)k c + R_\text{i}^2} }, \quad
	f_\text{sf}(R) = k \frac{ \sqrt{ ( R^2 - R_\text{i}^2  )/(k c) + r_\text{i}^2 } }{ R }.
\end{equation}

As is seen from the equations, geometrical parameters in both stress-free and load-free configurations uniquely define $\mathbf{F}_0$.
If some of the parameters $R_\text{i}$, $R_\text{o}$, $L$ are unknown, they can be obtained by solving an inverse problem.
Such a situation can occur, for example, when it is impossible to cut the artery to check its' stress-free kinematics.
Let us assume that $l$, $r_\text{i}$, $r_\text{o}$, $\alpha$ and material parameters of the specimen are known.
We need thus to find $L, R_\text{i}$ and $R_\text{o}$ in order to obtain $\mathbf{F}_0$.

Dealing with tubes made of an incompressible material, the entire kinematics of the body is a function of two scalar parameters.
This gives rise to the semi-analytical procedure described in Subsection 3.2 of \cite{Holzapfel2000} and in Appendix A of \cite{Tagiltsev2018}.
It allows one to calculate the internal pressure $p_\text{i}(L, R_\text{i}, R_\text{o})$ and the reduced axial force $F(L, R_\text{i}, R_\text{o}) = N - \pi r_\text{i}^2 p_\text{i}$ (where $N$ is a total axial force) which correspond to the deformation of the tube from its stress-free configuration.

For the inverse problem, the overall system of equations is as follows
\begin{equation} \label{SFSizes}
\begin{cases}
    \begin{array}{l}
        p_\text{i}(L, R_\text{i}, R_\text{o}) = 0 \\
        F(L, R_\text{i}, R_\text{o}) = 0
    \end{array}
%     \mid  \text{absence of loads condition}
     \\
    L (2\pi - \alpha)(R^{2}_\text{o} - R^{2}_\text{i}) = l \cdot 2\pi(r^{2}_\text{o} - r^{2}_\text{i}),
\end{cases}
\end{equation}
where the first two equations control the absence of loads in the lf-configuration and the third equation stands for the incompressibility of the material.
In the current study, system \eqref{SFSizes} is solved numerically using Newton-Raphson method.
Note that the incompressibility condition may be accounted for analytically, by taking $L = l \cdot k (r^{2}_\text{o} - r^{2}_\text{i})/(R^{2}_\text{o} - R^{2}_\text{i})$ in formulae for $\mathbf{F}_0$.
That leads to a reduced system with only two equations, which can be solved more efficiently.

\subsection{FEM simulation: cutting of an artery \label{FEM32}}

\begin{figure}\centering
	
    \scalebox{0.8}{
    \includegraphics{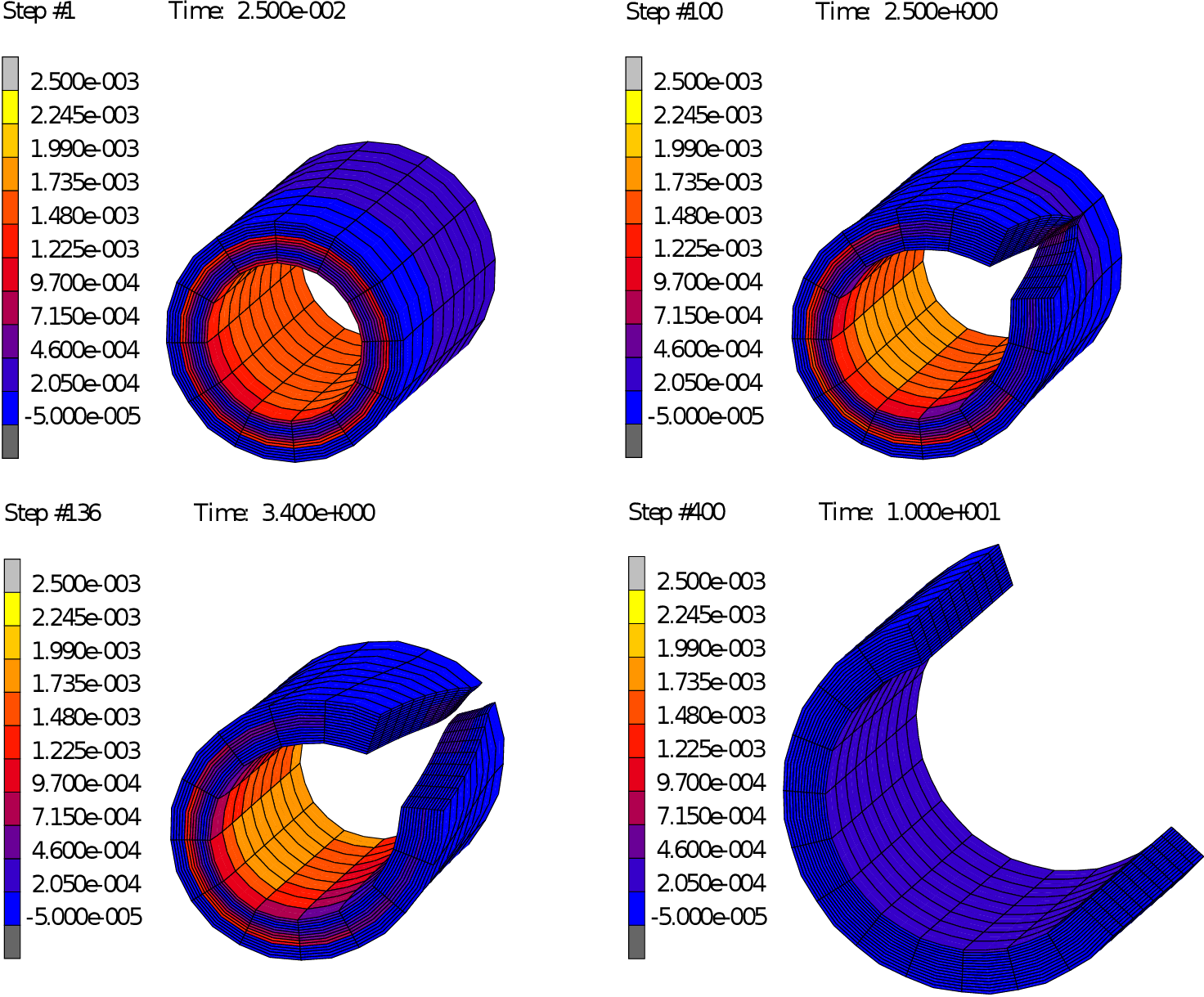}}
    \caption{FEM simulation of artery cutting with scalpel. \label{FEM32Fig}}
\end{figure}

To demonstrate the interrelation between approaches to residual stresses described above, FEM simulation of cutting of an artery with a scalpel is carried out.
The artery is considered as a double-layered viscoelastic composite tube.
The inner and outer layers correspond to media and adventitia, respectively.
Each layer is reinforced by two families of fibres.
Fibres are arranged in symmetrical helixes; they are inclined at the angles $\pm\beta$ to the hoop (circumferential) direction.
In the absence of viscous stresses (which corresponds to the quasi-static loading) the model reduces to the purely hyperelastic one, previously considered in \cite{Holzapfel2000} to describe the behaviour of an artery.
The material model of the viscoelastic composite is implemented into the commercial FEM code MSC.MARC using the Hypela2 interface for user-defined models.
Efficient stress computation algorithms proposed in \cite{ShutovLandgraf2013} and \cite{Tagiltsev2018} for the isotropic and fibre-like Maxwell bodies are used, respectively.
The inner layer (media) is discretized by 10 $\times$ 12 $\times$ 3 elements in the radial, circumferential and axial directions respectively.
The outer layer (adventitia) is subdivided into 5 $\times$ 12 $\times$ 3 elements.
The used elements are three-dimensional bricks of Hex20 type (twenty-node elements with a quadratic approximation of the geometry and displacements).
Herrmann formulation is employed, which includes one extra degree of freedom for pressure.
It allows one to model a nearly incompressible material behaviour in a consistent way thus preventing unphysical volumetric locking.
Dynamic transient problem statement is considered; to suppress undesired oscillations additional damping is applied.
The initial velocity field is equal to zero throughout the body.
Kinematic and material parameters of the simulation are listed in Tables \ref{FEMparamsKinem} and \ref{FEMparamsMater}, respectively.
The initial conditions for internal variables are set according to Subsection \ref{Initials}; the field $\mathbf{F}_0$ is pre-defined in the form \eqref{CalibrField}.
The simulation is performed for the total time $t_\text{total} = 10 s$ with the constant step size $\Delta t = 2.5 \cdot 10^{-2} s$, which requires 400 time steps.
The motion of the perfect scalpel is modelled by deliberate removing links between two contacting surfaces along the cut.
Deformed configurations and corresponding stress distributions at steps $\#1$, $\#100$, $\#136$ and $\#400$ are shown in Figure \ref{FEM32Fig}.
For the animation of the artery cutting, the reader is referred to \url{https://youtu.be/8v_2RtHuqUg}.
An animation of cutting the half of the artery along the hoop direction is available under \url{https://youtu.be/iQeY9UvtflI};
cutting of the outer layer (adventitia) in the hoop direction can be viewed under  \url{https://youtu.be/6FLOA-50e7k}.

\begin{table}
\caption{Kinematic parameters for the simulation in Section \ref{FEM32}}
\label{FEMparamsKinem}
\begin{center}
\begin{tabular}{c | c}
	\hline
	$r_\text{i}$, mm  & 0.71\\
	$r_\text{interface} $, mm & 0.97 \\
	$r_\text{o}$, mm & 1.1 \\
	$l$, mm & 3.0\\
	$R_\text{i}$, mm & 1.3948 \\
	$R_\text{interface}$, mm & 1.6589 \\
	$R_\text{o}$, mm & 1.8024\\
	$L$, mm & 2.9251\\
	$\alpha$, degrees & 160 \\
	
	\hline
\end{tabular}
\end{center}
\end{table}

\begin{table}
\caption{Material parameters for the simulation in Section \ref{FEM32}}
\label{FEMparamsMater}
\begin{center}
\begin{tabular}{c | c | c}
	Parameter & Media & Adventitia \\
	\hline
	$c_1$, KPa & 3.0 &  0.3 \\
	$c_2$, KPa & 2.0  & 0.2 \\
	$\eta_\text{matrix}$, KPa $\cdot$ s & $5 \cdot 10^{-1}$  & $1 \cdot 10^{-1}$\\
	$\mu_\text{matrix}$, KPa & $5 $ & $1 $\\
	$\beta$, degrees & 29 & 62\\
	$k_1$, KPa & 2.3632 & 0.562\\
	$k_2$ & 0.8393 & 0.7112\\
	$k_{1, \text{visc}}$, KPa & 5.3 & 1.3 \\
	$k_{2, \text{visc}}$ & 0.8393 & 0.7112 \\
	$\eta_\text{fibre},$ KPa $\cdot$ s & $5.3 \cdot 10^{-1}$ & $1.3 \cdot 10^{-1}$ \\
	\hline
\end{tabular}
\end{center}
\end{table}

It is worth emphasizing that in the simulation the artery behaves smooth from the start at the first simulation step $\#1, t = \Delta t $.
This indicates that the initial state is at equilibrium, as expected.
The final step of the simulation corresponds to the maximum opening angle, which is close to the pre-defined angle $\alpha = 160^{\circ}$.
As is seen, ``opening angle'' approach can be efficiently described by a more general $F_0$ approach.

%_____________________________________________________________________

\section{Case of different opening angles \label{DifAngles}}

%\subsection{Problem setting}

In a general case, different layers of a composite tube may experience different strains upon unloading.
Thus, the unloaded global configurations are, in general, incompatible: different layers in the load-free state do not fit together.
An extra load along the interface occurs if the layers are glued together to form a connected composite.
Therefore, the pre-stresses are defined by an interplay between the geometric mismatch of layers in the unstressed state and the corresponding material parameters.
The current section focuses on the influence of the geometry on residual stresses.

Let us consider a composite tube consisting of two layers.
We refer to them as media (M) and adventitia (A) like in the previous section.
In contrast to the previous section, now each layer \textit{has different opening angles},  which brings the layer to its stress-free configuration.
We denote the inner and outer radii, the length and the opening angle of the layers in the sf-configuration as $R_{\text{i},\text{type}}$, $R_{\text{o},\text{type}}$, $L_\text{type}$ and $\alpha_\text{type}$, respectively ($\text{type} \in \lbrace \text{M}, \text{A} \rbrace$).
In general, in their load-free states there is a mismatch between the layers.
The connected composite without external load is considered as the general load-free state of the composite tube.
Note that this lf-configuration does not have to coincide with any of individual lf-configurations of media and adventitia.
The inner and outer radii of the composite tube as well as the interface position are given by $r_\text{i}$, $r_\text{o}$, $r_\text{interface}$, respectively; the length equals $l$.
The following problem arises in a natural way: for known $R_{\text{i},\text{type}}$, $R_{\text{o},\text{type}}$, $L_\text{type}$, $\alpha_\text{type}$ and material parameters one needs to find $r_\text{i}$, $r_\text{o}$, $r_\text{interface}$, $l$ and to describe the mechanical behaviour of the pre-stressed composite.

\textbf{Remark 2.}
Note that this problem statement differs conceptually from the one considered in \cite{Grobbel2018} and \cite{Mousavi2017}. In contrast to our case, in these papers a  number of different unstressed configurations pertaining to different constituents are considered.

\begin{figure}\centering
    \psfrag{RoM}[m][][1][0]{$R_{\text{o},\text{M}}$}
    \psfrag{RiM}[m][][1][0]{$R_{\text{i},\text{M}}$}
    \psfrag{Am}[m][][1][0]{$2\pi-\alpha_\text{M}$}
    \psfrag{Lm}[m][][1][0]{$L_\text{M}$}
    \psfrag{RoA}[m][][1][0]{$R_{\text{o},\text{A}}$}
    \psfrag{RiA}[m][][1][0]{$R_{\text{i},\text{A}}$}
    \psfrag{Aa}[m][][1][0]{$2\pi-\alpha_\text{A}$}
    \psfrag{La}[m][][1][0]{$L_\text{A}$}
    \psfrag{Force}[m][][1][0]{$F$}
    \psfrag{l}[m][][1][0]{$l$}
    \psfrag{Press}[m][][1][0]{$p_{\text{i},\text{M}}$}
    \psfrag{Press2}[m][][1][0]{$p_{\text{o},\text{A}}$}
    \psfrag{Ri}[m][][1][0]{$r_\text{i}$}
    \psfrag{Rint}[m][][1][0]{$r_\text{interface}$}
    \psfrag{Ro}[m][][1][0]{$r_\text{o}$}
    \psfrag{Fa}[m][][1][0]{$\mathbf{F}_\text{A}$}
    \psfrag{Fm}[m][][1][0]{$\mathbf{F}_\text{M}$}
    \scalebox{1.0}{
    \includegraphics{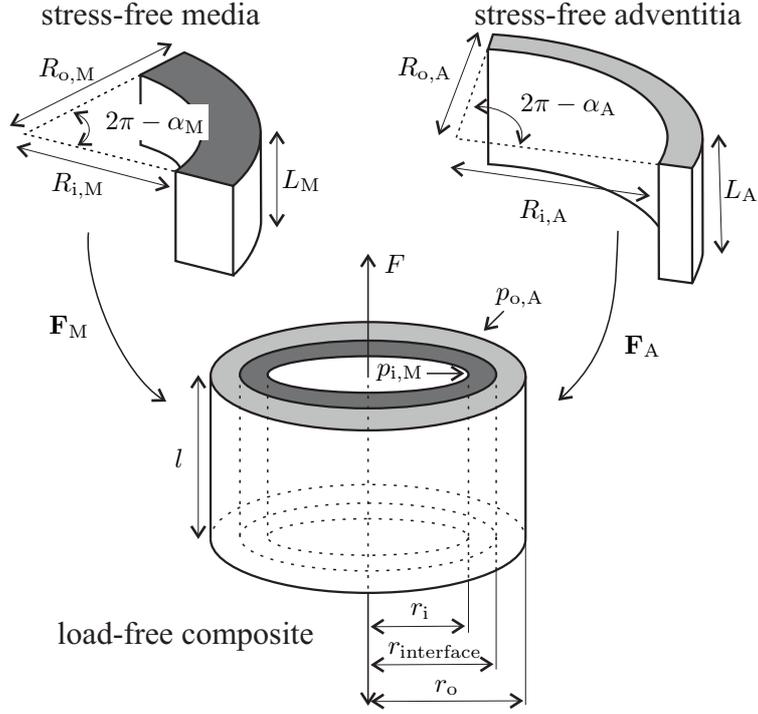}}
    \caption{Kinematics of two-layered composite \label{Scheme}}
\end{figure}

\subsection{Finding $r_\text{i}$, $r_\text{o}$, $r_\text{interface}$, $l$ \label{SemiAnalytDifAngles}}

Applying the semi-analytical procedure which was already mentioned in Section \ref{SemiAnalyt} to both layers we formulate the problem as follows.
In the absence of external loads pre-stressed composite tube remains in equilibrium state after two layers are glued together, see Figure \ref{Scheme}.
Denote the deformation of the media-layer from its stress-free configuration to the state of circular tube by $\mathbf{F}_\text{M}$; let $l$ and $r_\text{interface}$ be the length and the outer radius of the media-layer within the tube.
Analogically, $\mathbf{F}_\text{A}$ stands for the deformation of the adventitia-layer
from its sf-configuration to the tube with the length $l$ and the inner radius $r_\text{interface}$.
Given the incompressibility of the material, the inner radius of the media and the outer radius of the adventitia (which are inner and outer radii of the composite tube) are explicitly expressed as
\begin{equation}\label{radii}
	r_\text{i} = \sqrt{ r_\text{interface}^2 - \frac{L_\text{M}}{l} \frac{2 \pi - \alpha_\text{M}}{2 \pi}(R_{\text{o},\text{M}}^2 - R_{\text{i},\text{M}}^2 ) } , \quad
	r_\text{o} = \sqrt{ r_\text{interface}^2 + \frac{L_\text{A}}{l} \frac{2 \pi - \alpha_\text{A}}{2 \pi}(R_{\text{o},\text{A}}^2 - R_{\text{i},\text{A}}^2 ) }.
\end{equation}

The equilibrium of the composite tube means that after the total deformation of the layers from their sf-configurations the difference between the internal and external pressure should be zero:
\begin{equation}\label{FirstEq}
	p_\text{i,M} - p_\text{o,A} = \int_{r_{\text{i}}}^{r_{\text{o}}} \frac{\mathbf{T}_{\theta \theta} -  \mathbf{T}_{rr}}{r} dr =0.
\end{equation}
Moreover we require that the reduced axial force is zero as well:
\begin{equation}\label{SecondEq}
	F = \pi \int^{r_{o}}_{r_{i}}( 2\mathbf{T}_{zz} - \mathbf{T}_{\theta\theta} - \mathbf{T}_{rr}) r dr= 0.
\end{equation}
It equals the applied axial force minus the force exerted by the internal pressure on the sealed ends of the tube.

The system of equations \eqref{FirstEq},\eqref{SecondEq} is solved numerically with respect to unknown length of composite tube $l$ and the radius of contact between layers $r_\text{interface}$.
The reader interested in details is referred to \cite{Tagiltsev2018}.
%After the quantities $l$ and $r_{interface}$ (and hence $r_i$ and $r_o$ too, \eqref{radii}) are found, subsequent computations are possible.
The presented semi-analytical procedure is used to compute the geometrical
parameters $r_\text{i}$, $r_\text{interface}$, $r_\text{o}$, and $l$.
They are used to define pre-stressed state within general FEM simulations in the following subsection.

\subsection{Locking of composite layers \label{FEM42}}

Kinematic parameters of the model are found using the procedure described above and are listed in Table \ref{KinDif}.
For material parameters of each layer see Table \ref{MatDif}.
Knowledge of the kinematic parameters allows one to obtain $\mathbf{F}_0$ in the form \eqref{CalibrField}; it will be used both in semi-analytical and FEM computations.

\begin{table}
\caption{Kinematic parameters for the simulation \ref{FEM42}}
\label{KinDif}
\begin{center}
\begin{tabular}{c | c}
	\hline
	$R_\text{i,M}$, mm & 1.0 \\
	$R_\text{o,M}$, mm & 1.4\\
	$L_\text{M}$, mm & 1.0\\
	$\alpha_\text{M}$, degrees & 160 \\
	$R_\text{i,A}$, mm & 1.5 \\
	$R_\text{o,A}$, mm & 1.8\\
	$L_\text{A}$, mm & 1.0\\
	$\alpha_\text{A}$, degrees & 140 \\
	$r_\text{i}$, mm  & 0.4852\\
	$r_\text{interface} $, mm & 0.8749 \\
	$r_\text{o}$, mm & 1.1691 \\
	$l$, mm & 1.0063\\

	\hline
\end{tabular}
\end{center}
\end{table}

\begin{table}
\caption{Material parameters for the simulation \ref{FEM42}}
\label{MatDif}
\begin{center}
\begin{tabular}{c | c | c}
	Parameter & Media & Adventitia \\
	\hline
	$c_1$, KPa & 3.0 &  0.3 \\
	$c_2$, KPa & 2.0  & 0.2 \\
	$\eta_\text{matrix}$, KPa $\cdot$ s & $5$  & $1$\\
	$\mu_\text{matrix}$, KPa & $5 $ & $1 $\\
	$\beta$, degrees & 29 & 62\\
	$k_1$, KPa & 2.3632 & 0.562\\
	$k_2$ & 0.8393 & 0.7112\\
	$k_\text{1, visc}$, KPa & 5.3 & 1.3 \\
	$k_\text{2, visc}$ & 0.8393 & 0.7112 \\
	$\eta_\text{fibre},$ KPa $\cdot$ s & $5.3$ & $1.3$ \\
	\hline
\end{tabular}
\end{center}
\end{table}

\begin{figure}\centering
    \psfrag{E}[m][][1][0]{$\quad E, \mu J$}
    \psfrag{A}[m][][1][0]{$\tilde{\alpha}, degrees$}
    \psfrag{Am}[m][][1][0]{$\alpha_\text{M}$}
    \psfrag{Aa}[m][][1][0]{$\alpha_\text{A}$}
    \scalebox{1.0}{
    \includegraphics{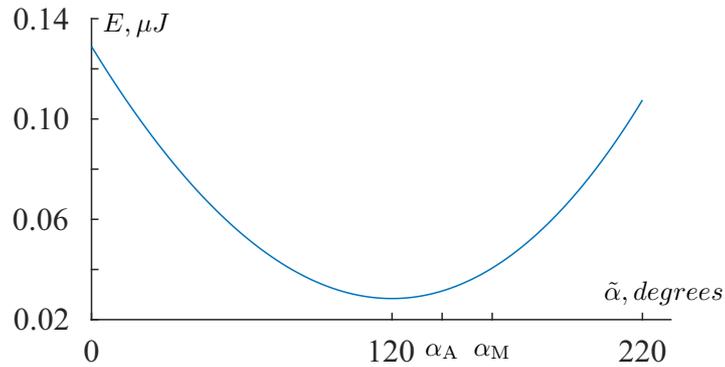}}
    \caption{ Dependence of the total free energy $E$ of the composite on the opening angle $\tilde{\alpha}$; the minimum of the free energy corresponds to the real opening angle of the composite. \label{EnergyCurve}}
\end{figure}

First, we solve the problem by the semi-analytical procedure.
Thus we neglect edge effects assuming that for each opening angle $\tilde{\alpha}$ the composite tube preserves its circular form.
The opening of the composite tube is described by the deformation gradient $\mathbf{F} = \text{diag}(\lambda_r, \lambda_\theta, \lambda_z)$, where $\lambda_{r}$, $\lambda_\theta$ and $ \lambda_z$ are material stretches in the radial, hoop and axial directions, respectively.
For each value of the angle $\tilde{\alpha}$, one can compute the overall stored energy  $E$ of the composite tube by integrating the Helmholz free energy through the tube's volume.
Naturally, the real opening angle $\alpha$ of the composite minimizes the energy:
\begin{equation}
	\alpha = \text{argmin} \;E(\tilde{\alpha}).
\end{equation}
The energy is plotted versus the angle $\tilde{\alpha}$ in Figure \ref{EnergyCurve}.
The optimum value of the angle is $\alpha = 120$ degrees.
This is much smaller than the opening angles of individual layers: $160$ degrees for the media and $140$ degrees for the adventitia.

To check this counterintuitive result, additional FEM computations are carried out.
In the FEM simulation the artery is cut along the axial direction at the very first time step (instant cut).
The type of the elements and the applied boundary conditions coincide with the ones from Section \ref{FEM32}; each layer is discretized now by 5 $\times$ 12 $\times$ 3 elements in the radial, hoop and axial directions, respectively.
The simulation is performed for the total time $t_\text{total} = 5 s$ with the constant step $ 5 \cdot 10^{-2} s$.

\begin{figure}\centering
    \scalebox{1.0}{
    \includegraphics{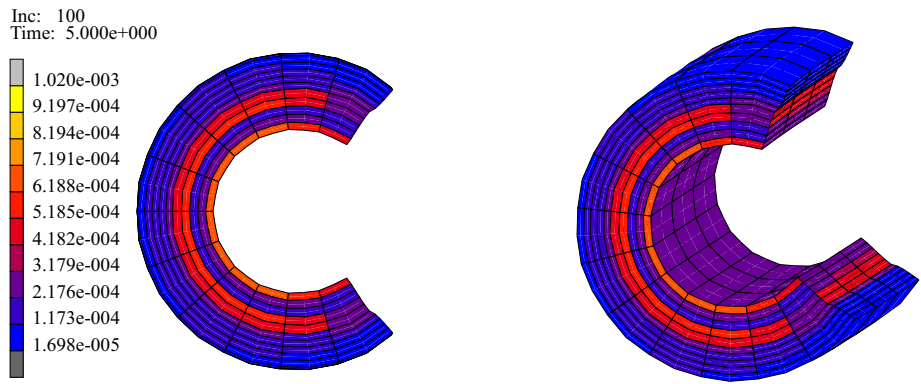}}
    \caption{Deformed configuration and corresponding stress distribution of the composite at the last time step, showing the mutual locking of layers. \label{FEM42Fig}}
\end{figure}

Deformed configuration and corresponding stress distribution of the composite for the last step are shown in Figure \ref{FEM42Fig} from two different viewing angles.
The last step is close to the asymptotic state where the body is at rest.
As is seen, stresses do not vanish completely in the final state.
The stresses remain after the cut since the opening angle of the composite tube is essentially smaller than the opening angles of both layers. Indeed, the final angle
 approximately equals $105 \div 115$ degrees, which is close
to the predictions of the semi-analytical procedure.
A minor mismatch between the semi-analytical
result and the FEM solution is caused by neglected edge effects.
Nevertheless, the discovered locking effect is not a consequence of the simplified kinematics adopted in the semi-analytical procedure.

To the best of our knowledge, the locking effect was not reported in the literature before.
We refer to this effect as to the \emph{mutual locking of layers}.
The locking appears due to the incompatible kinematics of unstressed layers.

%_____________________________________________________________________

\section{Conclusion \label{Concl}}

A geometrically exact approach to the modelling of viscoelastic composites is considered.
A special field $\mathbf{F}_0$ is used to account for the presence of residual stresses in the considered structure.
Given this field one can describe the kinematics of the material particle employing two different reference configurations (stress-free and load-free configurations are used in the current paper).
The applicability of the $\mathbf{F}_0$-approach to the multiplicative inelasticity (based on the Sidoroff decomposition of the deformation gradient) is demonstrated; the hyperelastic material behaviour is covered as a special case.
The $\mathbf{F}_0$-field can be interpreted as both inelastic and elastic deformations which appear in the multiplicative decomposition.
The positive feature of the $\mathbf{F}_0$-approach is that well-established numerical algorithms still can be used; the introduction of pre-stresses does not increase the complexity of the numerical schemes.

%A procedure for obtaining the residual stresses using the $\mathbf{F}_0$-approach is exemplified in terms of a fibre-reinforced viscoelastic composite structure which was advocated in \cite{Tagiltsev2018}; that model is based on the iso-strain assumption and uses multiplicative decomposition for the formulation of viscoelasticity.

The existence of pre-stresses is especially important in modelling of  biological tissues.
One of the first attempts to take the pre-stresses into account was the ``opening angle'' approach (used in \cite{Holzapfel2000}), which is based on the kinematic difference between stress-free and load-free configurations for the whole circular segment of an artery.
It is shown that the advocated $\mathbf{F}_0$-approach is capable of reproducing the ``opening angle'' approach as a special case.
Explicit expressions for the $\mathbf{F}_0$-field in such a case are provided.
As a demonstration of the interrelation between mentioned approaches, FEM simulations of cutting an artery are performed.
The number of material parameters which appear in the considered material model is relatively low and they possess a clear mechanical interpretation.
As a result, a simple modelling approach is obtained on the macro-scale which has a potential to become a practical clinical tool \cite{Nappi2016}.

In contrast to a wide-spread assumption that a sliced artery is stress free \cite{Chuong1986}, \cite{Fung1993}, \cite{Humphrey2004}, our simulation indicates that the stresses do not vanish completely.
This theoretical result is also confirmed by experiments reported in \cite{Greenwald1997}.
Moreover, semi-analytical and FEM solutions show the counterintuitive result that the opening angle of the composite tube is significantly smaller than the opening angles of each individual layer.
The discovered effect is referred to as the mutual locking of layers.
It is explained by incompatible kinematics upon unloading.
An important implication of this locking effect is that the cutting test by itself does not carry enough information about constituents of the composite.

An essential advantage of the $\mathbf{F}_0$-approach over the conventional ``opening angle'' approach lies in its flexibility.
In particular, different $\mathbf{F}_0$ fields can be used not only for different layers, but also for the different constituents like matrix and fibre.
Experimental findings indicate different opening angles for different constituents, like collagen fibres and myocites \cite{Grobbel2018}.
Also, the growth and remodelling of soft tissues can be taken into account by a proper evolution equation for $\mathbf{F}_0$.
In the follow-up studies the applicability of the $\mathbf{F}_0$-approach to complex biological structures will be investigated.

\textbf{Acknowledgements.}
The financial support provided by the RFBR (grant number 17-08-01020) and by
the integration project of SB RAS (project 0308-2018-0018) is acknowledged.

%_____________________________________________________________________

\end{document}